\documentclass[aps,prd,twocolumn,superscriptaddress,nofootinbib,showpacs,preprintnumbers]{revtex4}
\usepackage{amsfonts}
\usepackage{amsmath}
\usepackage{amssymb}
\usepackage{graphicx}
\usepackage{euscript}

\begin{document}

\title{FRW Quantum Cosmology with a Generalized Chaplygin Gas}~\thanks{This research work was supported by the grants
CERN/P/FIS/45459/2002 and FEDER-POCTI/P/FIS/32327/2000}

\author{Mariam Bouhmadi-L\'{o}pez}
\email{mariam.bouhmadi@port.ac.uk} \affiliation{Institute of
Cosmology and Gravitation, University of Portsmouth,  Mercantile
House, Hampshire Terrace,  Portsmouth  PO1 2EG, UK}

\author{Paulo Vargas Moniz}
\email{pmoniz@dfisica.ubi.pt}~\thanks{ URL: {\sf
http://www.dfis.ubi.pt/$\sim$pmoniz}}~~\thanks{Also at CENTRA,
IST, Av. Rovisco Pais, 1, 1049 Lisboa, Portugal}
\affiliation{Departamento de Fisica, Universidade da Beira
Interior, R. Marqu\^es d'Avila e Bolama,  6200 Covilh\~{a},
Portugal} \pacs{98.80.-k, 98.80.Qc \hfill PU-ICG-04/08,
gr-qc/0404111}

\begin{abstract}
Cosmologies with a Chaplygin gas have recently been explored with
the objective of explaining the transition from a dust dominated
epoch towards an accelerating expansion stage. We consider the
hypothesis that the transition to the accelerated period involves
a quantum mechanical process. Three physically admissible cases
are possible. In particular, we identify a minisuperspace
configuration with two Lorentzian sectors, separated by a
classically forbidden region. The Hartle-Hawking and Vilenkin wave
functions are computed, together with the  transition amplitudes
towards the accelerating epoch. Furthermore, it is found that for
specific initial conditions,
 the parameters characterizing the generalized Chaplygin gas
 become related through an expression involving an integer $n$.
We also introduce  a phenomenological association between some
brane-world scenarios and a FRW minisuperspace cosmology with a
generalized Chaplygin gas. The aim is to promote a discussion and
subsequent research on the quantum creation of brane cosmologies
from such a perspective. Results  suggest that the brane tension
would become related with generalized Chaplygin gas parameters
through another expression involving an integer.
\end{abstract}

\date{April 27, 2004}
\maketitle

\section{Introduction}

Mounting evidence from supernova Ia  (SNIa) observations
\cite{Riess:1998cb} and the more recent cosmic microwave
background radiation (CMBR) data (see, e.g.,
Ref.~\cite{Spergel:2003cb} and \cite{2a}) is suggesting  that the
expansion of our universe seems to be in an accelerated state.
This has been designated as the ``dark energy'' effect \cite{3a}.
A possible candidate responsible for this evolution is the usual
vacuum energy represented by a cosmological constant, $\Lambda $,
providing a negative pressure \cite{3b,3c}. However, this idea has
to deal with the fact that the observational value of $\Lambda$ is
$120$ orders of magnitude smaller than that established from field
theory methods \cite{3b,3c}. Moreover, it is a special ``cosmic
coincidence'' that the energy density associated with the
cosmological constant has a  value near to the density of other
matter types exactly today. A plausible alternative is to consider
a dynamical vacuum energy (also called ``quintessence'')
\cite{Wetterich:fm}, involving one or two scalar fields, where
some scenarios are associated with potentials justified from
supergravity theories \cite{Brax:1999yv}. Nevertheless, when
quintessence models confront the cosmic coincidence issue, they
face fine-tuning problems and no satisfactory solution has yet
been found.

Recently, an interesting proposal has been made \cite{4},
describing  a transition from a universe filled with dust-like
matter to an accelerating expanding stage: the scenario of the
Chaplygin gas applied to cosmology, which was later generalized in
Ref.~\cite{26,A}. The generalized Chaplygin gas model is described
by a perfect fluid obeying an exotic equation of state \cite{A}
\begin{equation}
p=-\frac{A}{\rho ^{\alpha }},  \label{cgi1}
\end{equation}
where $A$ is a positive constant and $0<\alpha \leq 1$. The
standard Chaplygin gas \cite{4} corresponds to $\alpha =1$.
Inserting this equation into the relativistic energy conservation
equation, leads to an energy density evolving as
\begin{equation}
\rho (a)=\left( A+\frac{B}{a^{3(1+\alpha )}}\right)
^{\frac{1}{1+\alpha }}, \label{cgi2}
\end{equation}
where $B$ is a constant. This model interpolates between a
universe dominated by dust and a DeSitter stage, through a phase
described by an equation of state of the form $p=\alpha \rho $. It
is curious to mention that the Chaplygin gas was originally
introduced in aerodynamics \cite{18a}.

Quite a few publications devoted to Chaplygin gas cosmological
models have already appeared in the literature
\cite{15,3rp,2rp,16,C,17,NewBD,18,20,21,23,27,Rev1,Rev2,DySy,Jackiw}.
Recent reviews can be found in \cite{Rev1,Rev2} and a dynamical
system analysis regarding the stability of FRW cosmologies with a
generalized Chaplygin gas was presented in Ref.~\cite{DySy}.

Within an observational context, it is worth  mentioning the
analysis of compatibility  at the level of CMBR \cite{2rp,3rp,15},
namely from recent WMAP data \cite{3rp}, together with SNIa
observations \cite{2rp} as well as the mass power spectrum of
large scale structure \cite{2rp,16} and unified dark matter
scenarios \cite{C}. Overall, some favourable indications have been
drawn when confronting generalized Chaplygin gas models with
cosmological observations.

Within a theoretical point of view, Chaplygin space-time models
have also  received increased attention. In particular, they can
be formally related to an effective description of a complex
scalar field whose action can be written as a generalized
Born-Infeld action \cite {A,26,8rc}, corresponding to a
``perturbed'' $d-$brane in a $(d+1,1)$ spacetime \cite{17}. In
particular conditions, a generalized Chaplygin gas can be obtained
in the context of a generalized Born-Infeld phantom framework
\cite{MBLJAJM}. For $\alpha =1$  the equation of state associated
with the parametrization invariant Nambu-Goto $d-$brane action in
a $(d+1,1)$ spacetime can be retrieved \cite{NewBD}. Moreover,
this action leads, in the light-cone gauge parametrization, to the
Galileo-invariant Chaplygin gas in a $(d,1)$ spacetime and to the
Poincar\'{e} invariant Born-Infeld action in a $(d,1)$ spacetime.
For an interesting survey relating fluid mechanics and $d$-branes
see also Ref.~\cite{Jackiw}.

The perspectives indicated in the previous paragraph
suggest that  the analysis of a space-time with a Chaplygin gas matter
content also conveys the opportunity to investigate brane-world physics
 phenomenologically.
Brane-worlds constitute a promising framework to explore
cosmological effects  (in view of the recent WMAP data about
inflation \cite{Spergel:2003cb}, as pointed out and discussed in
Ref.~\cite{lidseytavakol}), together with explaining other
fundamental physical issues (e.g., the hierarchy problem
\cite{ADD}). In the published literature the following scenarios
can be found regarding brane cosmologies: the Arkani-Hamed,
Dimoupolos and Dvali (ADD) model \cite{ADD}, the Randall-Sundrum
(RSI and RSII) models (see, e.g., Ref.~\cite{RS} and \cite{RSII},
respectively) and the ekpyrotic model \cite{ekpy}. For a recent
review on brane-worlds see Ref.~\cite{RMreview,BBD}. Pertinent issues
concerning brane-world cosmologies include \cite{bilbao}: ($i$)
Early evolution of the universe, ($ii$) perturbations on the
brane/bulk and spectrum of fluctuations, ($iii$) stability of
extra dimensions, ($iv$) onset of inflation, ($v$) dark
matter/energy problems, ($vi$) corrections to Einstein gravity and
($vii$) novel issues in geometry (e.g., non-commutative
geometries). Besides the previous topics, brane-world cosmology
also has to address the fundamental question of the origin of our
universe. Interesting proposals, regarding the quantum
cosmological creation of brane world models, were recently
introduced and analysed in
Ref.~\cite{ekpy,israel,leon,odintsov,j,k,l,m}.

%%%%%%%%%%%%%%%%%%%%%%%%%%%%%%%%%%%%%%%%%%%%%%%%%%%%%%%%%%%%%%%%%%%%%%%%
%%%%%%%%%%%%%%%%%%%%%%%%%%%%%%%%%%%%%%%%%%%%%%%%%%%%%%%%%%%%%%%%%%%%%%%%%
%%%%%%%%%%%%%%%%%%%%%%%%%%%%%%%%%%%%%%%%%%%%%%%%%%%%%%%%%%%%%%%%%%%%%%%%

Within the physical  context herein described, the main purpose of
this paper is to investigate if a quantum mechanical analysis
would provide  complementary information regarding FRW cosmologies
with a Chaplygin gas content. In particular, assuming that after
the universe reached a stage where dust matter was dominant,  the
subsequent evolution towards  an exponentially expanding epoch
involved a quantum mechanical transition associated with some
remnant component of the original wave function of the universe.
In other words,  underlying the classically
observed universe on large scales there is a quantum mechanical
background, where the cosmological influence of a wave function vestige (i.e., a rapidly
oscillating state with a particularly small amplitude)
would emerge during the dust dominace epoch.
This specific wave function remnant had been present
through that stage of the  universe evolution and could be considered as a
robust component with respect to decoherence processes in the
early universe (see Ref.~\cite{CKZeh,Decoh,Diosi} for details).
In this framework, some pertinent questions may be raised.
Could such a quantum
 description determine important physical consequences
regarding a cosmological Chaplygin gas scenario? Namely, determining
the transition probabilities towards an accelerated stage and restricting the
range of values for the Chaplygin gas parameters?

We  also focused on an additional aim: a discussion on
the  quantum  creation of a
brane-world cosmology. This perspective is suggested by
 the phenomenological context that some brane-worlds acquire
within generalized Chaplygin gas cosmologies. This allows us to
make use of  important similarities between the quantum cosmology
of a brane-world model and a FRW quantum cosmology with a
Chaplygin gas content, which we will point in detail herein. While
the former may require some non-standard tools (see e.g.,
Ref.~\cite{israel}), the latter imports techniques from well known
quantum cosmological developments \cite{CKbook,qc,dw,ck}. In more
detail, an analysis can be made exploring  the minisuperspace
resemblances between particular brane-worlds, generalized
Chaplygin gas cosmological models and   FRW minisuperspace
scenarios with a complex scalar field $\phi $, with $\phi
  \equiv e^{i\theta }\varphi $ (where $\theta$ constitutes a cyclical coordinate,
  determining that the conjugate momentum
  $\pi_\theta $ corresponds to  a conserved
charge). We will present herein some interesting results which are
retrieved when including additional
quantum corrections to the WKB description.

 Accordingly, this paper is organized as follows.
In Sec.~II we present our minisuperspace model, pointing
to the possible relation between a generalized Chaplygin gas cosmology and some
brane-world properties.  The
corresponding Hartle-Hawking and Vilenkin states are obtained in
Sec.~III. Interesting quantization consequences are described in
Sec.~IV, and possible extensions beyond the saddle-point
approximation are discussed in Sec.~V. Sec.~VI constitutes a
summary of the results herein presented. An appendix is also
included. It conveys  the calculations of turning points, which
are pertinent for  the  minisuperspace case (investigated
throughout sections II to V), whose interesting similarities with
some brane-world cosmologies are discussed in this paper.

\section{Minisuperspace Model}

Let us  take the case of a closed ($k=1$) FRW model in the
presence of a positive cosmological constant $\Lambda >0$ with a
generalized Chaplygin gas\footnote{See Sec.~VI for a discussion on
the presence of $\Lambda$: it makes our model analytically
feasible and can represent the lowest order effect of the
effective cosmological constant determined by the generalized
Chaplygin gas.}. The minisuperspace Lagrangian has the form (see,
e.g., ref. \cite{HE,IRAN})
\begin{equation}
L=-\frac{3\pi }{4G}\left( \frac{\dot{a}^{2}a}{N}-Na\right) -2\pi a^{3}N\frac{%
\Lambda }{8\pi G}-2\pi a^{3}N\rho .  \label{cg1}
\end{equation}
After determining the corresponding canonical momentum to $a$ and
therefore the Hamiltonian constraint,
\begin{equation}
 -\frac G{3\pi}\pi_a^2 +\frac{3\pi}{4G}V(a)=0, \label{cgnew01}
\end{equation}
its quantum description can be represented by the Wheeler-DeWitt
equation, written as\footnote{We are ignoring factor ordering
aspects related with $\pi^2_a$, with $\pi_a$ being the conjugate momentum to $a$.}
\begin{equation}
\left[ -\frac G{3\pi}\frac{d^{2}}{da^{2}}+\frac{3\pi}{4G}V(a)\right]
\Psi(a)=0,  \label{cg2}
\end{equation}
where $\Psi(a)$ is the wave function characterizing the quantum state of
this FRW universe. The effective minisuperspace potential $V(a)$ is given by
\begin{equation}
V(a)=a^{2}-\lambda a^{4}-\bar\rho(a)a^{4},  \label{cg3}
\end{equation}
where $\lambda=\frac\Lambda3$,
\begin{equation}
\bar\rho(a)=\left( \bar{A}+\frac{\bar{B}}{a^{3(1+\alpha)}}\right) ^{\frac
1{1+\alpha}},  \label{cg4}
\end{equation}
together with $0<\alpha\leq1$, $A=\bar{A}\left( \frac{8\pi G}3\right)
^{\alpha+1}$, $B=\bar{B}\left( \frac{8\pi G}3\right) ^{\alpha+1}$, $A,B\in%
\mathbb{Re}^{+}$. As mentioned, for $\alpha=1$ we recover the
usual Chaplygin gas case. We will consider a  semiclassical WKB
(Wentzel, Kramers, Brillouin) description, where the wave function
of the universe can be approximated by
\begin{equation}
\Psi=C_1\psi_{1}+C_2\psi_{2},  \label{cg5}
\end{equation}
where
\begin{equation}
\psi_{i}=\exp\left[ -\frac1GS_{0}^{(i)}(a)\right],  \label{cg6}
\end{equation}
with $i=1,2$ and $S_{0}^{(i)}(a)$ satisfying the Hamilton-Jacobi equation
\begin{equation}
\left( \frac{dS_{0}^{(i)}(a)}{da}\right) ^{2}=\frac{9\pi^{2}}4V(a).
\label{cg7}
\end{equation}
In particular,
 $\psi_{1}$ and $\psi_{2}$  constitute outgoing or ingoing
modes in the classically allowed regions,  decreasing or
increasing functions in the classically forbidden regions,
respectively,  and $C_1,C_2$ are constants.
The WKB validity condition is given by $G\left|
\frac{dV(a)}{da}\right| \ll\left| V(a)\right| ^{3/2}.$

In order to retrieve the wave functions in (\ref{cg5}) in an
analytical form, a saddle-point formulation will be employed.
Therefore, it is necessary to determine the possible turning
points for the potential $V(a)$. Analyticity also requires we
employ the following approximation (where we henceforth drop the
overbars  on $A$ and $B$)
\begin{eqnarray}  \label{cg8}
\left( A+\frac B{a^{3(1+\alpha)}}\right)
^{\frac1{1+\alpha}}\approx
\frac{B^{\frac1{1+\alpha}}}{a^{3}}\left[ 1+\frac1{1+\alpha}\frac
ABa^{3(\alpha+1)}+\right.\nonumber\\
\left.\frac12\frac1{1+\alpha}\left( \frac1{1+\alpha}-1\right)
\frac{A^{2}}{B^{2}}a^{6(\alpha+1)}+\ldots\right]. \nonumber \\
\end{eqnarray}
and if we take
\begin{equation}
a\ll\left[ \left( \alpha+1\right) \frac BA\right] ^{\frac1{3(\alpha+1)}},
\label{cg9}
\end{equation}
we can write the potential $V(a)$ as
\begin{equation}
V_{0}(a)\approx a^{2}-\lambda a^{4}-aB^{\frac1{1+\alpha}},
\label{cg10}
\end{equation}
where we neglected the second as well as higher order terms in
(\ref{cg8}) as $0<\alpha\leq1$ (see Eq.~(\ref{cg3}) and
(\ref{cg8})). We can thus identify a suitable range of values for
the parameters $A$, $B$, $\alpha$, and the scale factor $a$ where
Eq.~(\ref{cg10}) is valid, corresponding to the analysis presented
in this paper. As far as the next order of approximation is
concerned, we will point in section V how the term proportional to
$a^{4+3\alpha}$ in $V(a)$ can be dealt with in a perturbative
manner when $A\ll1$.

In the linear approximation provided by Eq.~(\ref{cg10}) we can
identify three physically different cases \cite{X,W}. A plot for
$V_{0}(a)$ regarding the second and third cases is given in
Fig.~\ref{potenciales} (see appendix A for more details):

\begin{figure}[h]
\includegraphics[width=\columnwidth]{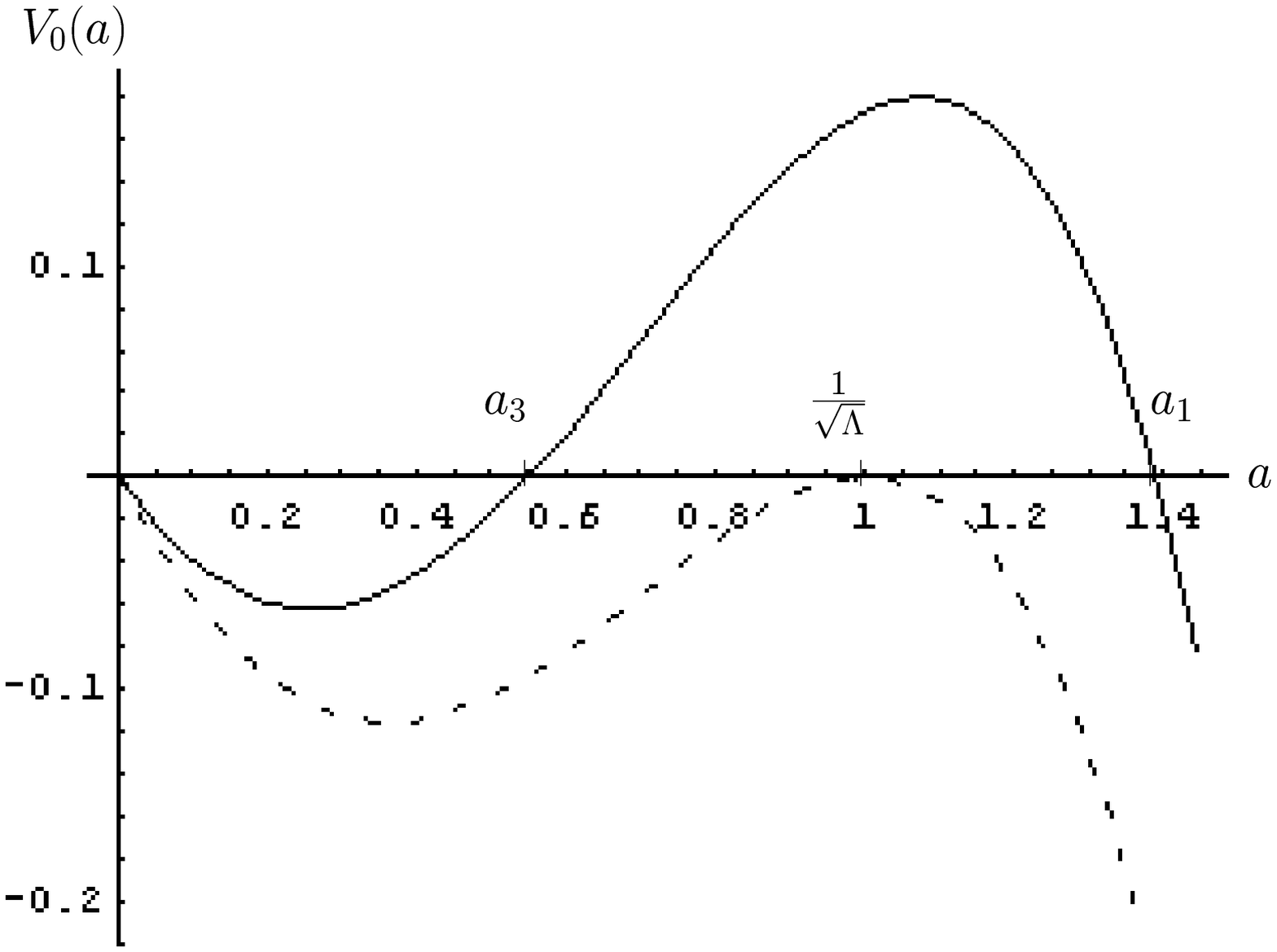}
\caption {This figure shows the behavior of the potential $V_0(a)$
 as function of $a$. The
dashed line corresponds to the case $\Lambda \tilde{B}^2=4/9$. In
this case, $V_0(a)$ vanishes at $a=1/\sqrt{\Lambda}$ and there are
two Lorentzian zones corresponding to $a<1/\sqrt{\Lambda}$ and
$1/\sqrt{\Lambda}<a$. The solid line corresponds to $\Lambda
\tilde{B}^2<4/9$. The potential vanishes at $a=a_1$ and $a=a_3$.
In this case there are two Lorentzian zones ($a<a_3$ and $a_1<a$)
separated by a classically forbidden region
($a_3<a<a_1$).}\label{potenciales}
\end{figure}

\begin{enumerate}

\item When $\Lambda\tilde{B}%
^{2}>\frac49$, where $\tilde{B}\equiv B^{\frac1{1+\alpha}}$, there is one
real root and pair of complex conjugated roots for the equation $%
f(a)=a^{3}-a/\lambda+\tilde {B}/\lambda=0$. The only value for which $%
V_{0}(a)$ vanishes is $a=0$ and $V_{0}(a)<0$ for $a>0.$

\item In this case,
we have $\tilde{B}=\frac 23\frac1{\sqrt{\Lambda}}$. It is found that $%
V_{0}(a)$ vanishes for $a=0$ and there is a positive root for $f(a)=0$ when $%
a=1/\sqrt{\Lambda}$. The other root is at $a=-2/\sqrt{\Lambda}$. We further
have that $%
V_{0}(a)<0$ for $0<a<1/\sqrt {\Lambda}$ and $a>1/\sqrt{\Lambda}.$

\item In the final  case, we have
$\Lambda\tilde{B}^{2}<\frac49$ and three real roots can be found
such that two are positive: $a_{1}>a_{3}>0>a_{2}.$
Moreover, $V_{0}(a)<0$ for $0<a<a_{3}$ and $a>a_{1}$, but $V_{0}(a)>0 $ for $%
a_{3}<a<a_{1}$. We thus have a classically allowed region followed by a
classically forbidden Euclidean region and finally there is a second
classically allowed Lorentzian region. For $0<a<a_{3}$ there is always a
region such that the linear approximation (\ref{cg10}) is valid. For $%
a_{3}<a<a_{1}$ this approximation is valid for $a_{1}\ll a_{\max}$ where $%
a_{\max}=[(\alpha+1)\frac B A]^{\frac 1{3(\alpha+1)}}$. For $a>a_{1}$, if $%
a_{1}\ll a_{\max}$, the linear approximation is valid if there are values of
$a$ such that $a_{1}<a\ll a_{\max}$. This leads to $\left( \frac 2{\sqrt{%
\Lambda}}\right) ^{3(\alpha+1)}\ll\frac BA$, which together with $\Lambda
B^{\frac2{1+\alpha}}<\frac49$ implies that $3^{3\left( \alpha+1\right)
}AB^{2}\ll1.$

\end{enumerate}

Before proceeding to obtain quantum solutions for the
Wheeler-DeWitt equation (see Eq.~(\ref{cg2}) and (\ref{cg10})), it
is of relevance to point the following. Quantum cosmological
scenarios with a physical configuration similar to the third case
 above described (where $\Lambda\tilde{B}^{2}<\frac49$ ), can be
found in the literature but in somewhat differently related
contexts.

One such situation is present in Ref.~\cite{5rp,6rp}
when  a FRW minisuperspace with a complex scalar field
of the form $\phi=e^{i\theta}\varphi$, with a mass term
proportional to $\varphi^{2}$ and a conserved charge (i.e.,
conjugate momentum) associated with $\pi_{\theta}$ is considered.
This produces
a minisuperspace potential $U(a,\varphi;\pi_{\theta})$ where
$\pi_{\theta}$ is a constant parameter. For arbitrary constant
values of $\varphi$ we obtain different sections in the potential
$U$, inducing  a similar dependence on $a$ as in $V_{0}(a)$ given
by Eq.~(\ref{cg10}) and depicted in Fig.~\ref{potenciales}.
Generalized tunnelling (Vilenkin) \cite{Tunnelling} and
Hartle-Hawking \cite{Hartle-Hawking} boundary conditions were then
introduced in Ref.~\cite{5rp,6rp}. The reason that the physical
model in Ref.~\cite{5rp,6rp} relates to the above Chaplygin gas
Hamiltonian formulation could be traced to the fact that the
Chaplygin gas FRW action can be retrieved from the action of FRW
models with the complex scalar field $\phi=e^{i\theta}\varphi$,
under specific conditions \cite{A,4,26}.

The association between  generalized Chaplygin gas cosmologies and
minisuperspace models with complex scalar fields can be brought
towards a recent and promising  domain: that of  brane cosmology
\cite{RMreview,BBD}. In fact,  the complex scalar field employed
in Ref.~\cite{A,26} to obtain a generalized Chaplygin gas can be
identified within a Born-Infeld action, which arise form the
imbedding of the (3+1)-D brane into the (4+1)-D spacetime
\cite{17}. Moreover, it can also be seen to correspond to a
``perturbed'' $d-$brane in a $(d+1,1)$ spacetime \cite{17}.
Furthermore actions of the Born-Infeld have recently been the
subject of wide interest (see ref. \cite{8rc} and references
therein). This comes comes from the result that the effective
action for the open string ending on $d-$branes can be written in
a Born-Infeld form. In that respect, Born-Infeld cosmological
solutions could assist us in the understanding of some brane-world
dynamics regarding the universe evolution.

The above context regarding brane-worlds can be further specified.
In fact,  FRW models with a Chaplygin gas may constitute a
phenomelogical tool to study  some brane-world models: another
physical situation with two Lorentzian sectors separated by a
classically forbidden Euclidean region can be found in the
Hamiltonian treatment of a DeSitter $(3+1)-$brane imbedded in a
$(4+1)-$Minkowski background \cite{israel}. The presence of a
classically disconnected epoch  (as in Fig.~1 for  $a<a_3$) has
been identified as a property of brane-world quantum cosmology
\cite{israel}. These remarks thus suggest the view that FRW models
with a generalized Chaplygin gas content would enable us   to
investigate  specific aspects of particular brane-world models. In
particular  through an Hamiltonian formulation, where
 symmetries and conserved quantities, possibly related with physical properties of
brane cosmologies, will be of relevance.
 In Sec.~V, where we consider
the next order of approximation in Eq.~(\ref{cg2}) and
(\ref{cg10}) (going beyond the Hamilton-Jacobi equation
(\ref{cg7})), we will point a further correspondence between
brane-world parameters and a generalized Chaplygin gas.

\section{Hartle-Hawking and Vilenkin Wave Functions}

We begin by addressing the  second
case $\left( \tilde{B}=\frac23\frac1{\sqrt{\Lambda}%
}\right) $:  we have two classically allowed Lorentzian regions
separated by the point $a=1/\sqrt{\Lambda}$ where $V_{0}(a)$
vanishes. Let us first analyze the region where
$a>1/\sqrt{\Lambda}$. Besides having to satisfy the range present
in Eq.~(\ref{cg9}) it is also necessary to have
\begin{equation}
\frac1{\sqrt{\Lambda}}\ll\left[ \left( \alpha+1\right) \frac BA\right]
^{\frac1{3(\alpha+1)}},  \label{cg15}
\end{equation}
(in order to have values of $a$ such that the linear approximation is still valid)
which leads to
\begin{equation}
B^{4}\ll A\left( \frac23\right) ^{\alpha+1}.  \label{cg16}
\end{equation}
The wave function is then expressed in the form of Eq.~(\ref{cg5})
and (\ref {cg6}) where
\begin{equation}
S_{0}(a)=i\frac{3\pi}2\int_{\frac1{\sqrt{\Lambda}}}^{a}\sqrt{-V_{0}(a)}da.
\label{cg17}
\end{equation}
Introducing the Lorentzian conformal time $\eta$ using
\begin{equation}
\frac{dS_{0}}{da}=i\frac{3\pi}2\frac{da}{d\eta},  \label{cg18}
\end{equation}
we get the following equation for $a(\eta)$
\begin{equation}
\frac{da}{d\eta}=\sqrt{-V_{0}(a)},  \label{cg19}
\end{equation}
from where it can be found that
\begin{equation}
a(\eta)=\frac1{\sqrt{\Lambda}}\left( \frac3{\cosh(\eta)-2}+1\right),
\label{cg20}
\end{equation}
with $\eta\in\left] -\infty,-\cosh^{-1}(2)\right[ $, noticing that
when $ \eta\rightarrow-\infty$ then
$a\rightarrow\frac1{\sqrt{\Lambda}}$ and when $
\eta\rightarrow-\cosh^{-1}(2)$ then $a\rightarrow+\infty$. For the
wave function it is obtained that
\begin{align}
\int_{\frac1{\sqrt{\Lambda}}}^{a}\sqrt{-V_{0}(a)}da&
=\frac13\left[ a\left( a+\frac2{\sqrt{\Lambda}}\right) \right]
^{\frac32}\nonumber\\
& -\frac 1{\sqrt{\Lambda} }\left( a+\frac1{\sqrt{\Lambda}}\right)
\sqrt{a\left( a+\frac2{\sqrt{\Lambda}
}\right) }  \nonumber \\
& +\frac1{\Lambda^{\frac32}}\ln\left[ \sqrt{a\left(
a+\frac2{\sqrt{\Lambda} }\right) }+\left(
a-\frac1{\sqrt{\Lambda}}\right) \right] \nonumber\\ &+\frac{\sqrt
{3}-\ln\left[ \left( 2+\sqrt{3}\right) /\sqrt{\Lambda}\right] }{
\Lambda^{\frac32}}\,\, .  \label{cg21}
\end{align}

Employing the same method for the region where
$a<1/\sqrt{\Lambda}$ it is found \cite{W,X}  that
\begin{equation}
a(\eta)=\frac1{\sqrt{\Lambda}}\left( -\frac3{\cosh(\eta)+2}+1\right),
\label{cg22}
\end{equation}
with $\eta\in\left[ 0,+\infty\right[ $, noticing that when $\eta
\rightarrow+\infty$ then $a\rightarrow\frac1{\sqrt{\Lambda}}$ and
when $\eta\rightarrow0$ then $a\rightarrow0$. For the wave
function it can be concluded that
$\int_{a}^{\frac1{\sqrt{\Lambda}}}\sqrt{-V_{0}(a)}da$ is equal to
minus the expression in Eq.~(\ref{cg21}).

In the next paragraphs we will retrieve
for the third case
$\Lambda\tilde{B}^{2}<\frac49$ the Hartle-Hawking
 \cite{Hartle-Hawking} and tunneling
(Vilenkin) \cite{Tunnelling} wave functions (see also
Ref.~\cite{CKbook,qc,dw,ck} for more details). In order  to obtain
the solutions $S_{0}(a)$ of Eq.~(\ref{cg7}) it is necessary to
determine the analytical expressions
\begin{equation}
\int\sqrt{\left| V_{0}(a)\right| da}\equiv\sqrt{\frac\Lambda3}\int I(a)da.
\label{cg11}
\end{equation}
This integral will of course depend on the specific chosen region, e.g.,
between $0<a<a_{3}$ for the first classically allowed Lorentzian region.
Let us determine the values for the integral in (\ref{cg11}) in
the following. After some lengthy calculations, it is found that \cite{W,X}
\begin{eqnarray}
\int_{a_{1}}^{a}I(a)da &=&\frac{aI}3-\beta\frac{a^{2}(a-a_{1})}
I\nonumber\\
&+& \gamma\left\{ a_{1}a_{3}(a_{1}^{2}-a_{3}^{2})
\Pi\left( \upsilon ,\frac{
2a_{1}+a_{3}}{2a_{3}+a_1},q\right) \right.\nonumber \\
&+& \big[a_{1}a_{3}^{2}(a_{1}+a_{3})-6\beta^{2}+\beta
a_{3}^{2}\big]F\left(
\upsilon,q\right) \nonumber\\
&+&\left.\beta\frac{\left( a_{1}+a_{3}\right)
^{2}}{2a_{3}+a_{1}}\Big[\left( a_{3}-a_{1}\right) \Pi\left(
\upsilon,1,q\right)\right.\nonumber\\
&+& \left.\left( 2a_{1}+a_{3}\right) F\left(
\upsilon,q\right)\Big] \right\}, \label{cg12}
\end{eqnarray}
where $\beta\equiv\frac{\left(
a_{1}^{2}+a_{1}a_{3}+a_{3}^{2}\right) }3$, $ \gamma\equiv\left(
\sqrt{a_{1}(2a_{3}+a_{1})}\right) ^{-1}$, $q\equiv \sqrt{
\frac{a_{3}\left( 2a_{1}+a_{3}\right) }{a_{1}\left(
2a_{3}+a_{1}\right) }}$ , $\upsilon\equiv\arcsin\sqrt{\frac{\left(
2a_{3}+a_{1}\right) \left( a-a_{1}\right) }{\left(
2a_{1}+a_{3}\right) \left( a-a_{3}\right) }}$, together with

\begin{eqnarray}
\int_{a}^{a_{3}}I(a)da&=&-\frac{aI}3-\beta\frac{a^{2}(a_{3}-a)}
I\nonumber\\ &+& \gamma\Big\{ a_{1}a_{3}\left(
a_{1}+a_{3}\right)\Big.\nonumber\\ &\times& \Big.\left[ \left(
a_{3}-a_{1}\right) \Pi\left( \chi,\frac{a_{3}}{a_{1}},q\right)
+a_{1}F\left(
\chi,q\right) \right] \Big.  \nonumber \\
&+& \beta\frac{\left( a_{1}+a_{3}\right) ^{2}}{2a_{1}+a_{3}}
\nonumber\\ &\times& \Big[ \left( a_{1}-a_{3}\right) \Pi\left(
\chi,q^{2},q\right) +\left( 2a_{3}+a_{1}\right)
F\left( \chi,q\right) \Big]  \nonumber \\
&-&\Big.6\beta^{2}F\left( \chi,q\right) +\beta\gamma
a_{1}^{2}F\left( \chi,q\right) \Big\}, \label{cg13}
\end{eqnarray}
 with $\chi\equiv\arcsin\sqrt{\frac{a_{1}\left(
a_{3}-a\right) }{ a_{3}(a_{1}-a)}}$, as well as
$\int_{a_{3}}^{a_{1}}I(a)da=\lim_{a\rightarrow
a_{1}}\int_{a_{3}}^{a}I(a)da$ where
\begin{eqnarray}
\int_{a_{3}}^{a}I(a)da &=& \frac{aI}3+\beta\frac{a^{2}(a-a_{3})}
I\nonumber\\&+&\gamma\left\{ -a_{1}a_{3}^{2}\left(
a_{1}+a_{3}\right) \Pi\left( \delta,
\frac{a_{1}-a_{3}}{a_{1}},r\right) \right.  \nonumber \\
&+&6\beta^{2}F\left( \delta,r\right)\nonumber\\
&-&\beta a_{1}\Big[ \left( a_{1}-a_{3}\right) F\left(
\delta,r\right) +a_{3}\Pi\left( \delta,1,r\right)
\Big] \nonumber \\
& +& \left. \beta\left( a_{1}+a_{3}\right) \right.\nonumber\\
&\times& \left.\Big[ a_{3}\Pi\left( \delta,r^{2},r\right) -\left(
2a_{3}+a_{1}\right) F\left( \delta,r\right) \Big]
\right\},\nonumber\\ \label{cg14}
\end{eqnarray}
with $\delta\equiv\arcsin\sqrt{\frac{a_{1}\left( a-a_{3}\right)
}{\left( a_{1}-a_{3}\right) a}}$ and
$r\equiv\sqrt{\frac{a_{1}^{2}-a_{3}^{2}}{\left(
2a_{3}+a_{1}\right) a_{1}}}$. The functions $F$ and $\Pi$ are
elliptic  integrals of the first and third kind, respectively
(see, e.g., Ref.~\cite{W,X} for more details). When calculating
the expression $\lim_{a\rightarrow a_{1}}\int_{a_{3}}^{a}I(a)da$
it is found that it is well defined except for the terms
corresponding to $\frac{a^{2}\left( a-a_{3}\right) }I+\frac
12\int_{a_{3}}^{a}\frac{a_{1}^{2}\left( a_{1}-a_{3}\right)
}{\left(
a-a_{1}\right) I}da$, as each of these expressions diverges when $%
a\rightarrow a_{1}$. However, it can be seen that these
divergences cancel each other \cite{W} and therefore the integral
in (\ref{cg14}) is well defined and can be used to determine the
transition amplitudes for the FRW universe with a generalized
Chaplygin gas to tunnel from a Lorentzian region between
$0<a<a_{3}$ to another where $a>a_{1}$ (see Fig.~
\ref{potenciales}).

Following Ref.~\cite{5rp,6rp} the no-boundary (Hartle-Hawhing)
wave function representing the region under the potential barrier
$V_{0}(a)$ corresponds to the increasing modes

\begin{equation}
\Psi\propto\exp\left[\frac{3\pi}{2G}\int_{a_{3}}^{a\leq
a_{1}}\sqrt {V_{0}(a)}da\right],\label{cg23}
\end{equation}
where we are using $V_{0}(a)\sim-\lambda
a(a-a_{1})(a-a_{3})(a+a_{1}+a_{3})$ . In more precise terms we
obtain \cite{29}
\begin{align}
\Psi(a) & = C\cos\left[ \frac{3\pi}{
2G}\int_{a}^{a_{3}}\sqrt{-V_{0}(a)}
da+\frac\pi4\right] ,\   (a<a_{3})&&  \label{cg24a} \\
\Psi(a) & =C\exp\left[ \frac {3\pi}{
2G}\int_{a_{3}}^{a}\sqrt{V_{0}(a)}
da\right] ,\   (a_{3}<a<a_{1})&&  \label{cg24b} \\
\Psi(a) & =2\tilde C\cos\left[
\frac{3\pi}{2G}\int_{a_{1}}^{a}\sqrt{-V_{0}(a)
}da-\frac\pi4\right] ,\  (a_{1}<a),&&  \label{cg24c}
\end{align}
where $\tilde{C}\equiv C\exp\left[
\frac{3\pi}{2G}\int_{a_{3}}^{a_{1}}\sqrt{V_{0}(a)}da\right] $ and
$C$ is an arbitrary constant.

As far as the tunnelling (Vilenkin) wave function is concerned,
it contains only outgoing modes and  we find

\begin{eqnarray}
\Psi(a)&=& C \left\{\exp\left[ -\frac{3\pi
}{2G}\int_{a_{3}}^{a_{1}}\sqrt{V_{0}(a)} da\right]\right.\nonumber\\
&\times& \left.\cos\left[ \frac{3\pi}{2G}
\int_{a}^{a_{3}}\sqrt{-V_{0}(a)}
da+\frac\pi4\right]\right.  \nonumber  \\
& + &\left.4i \exp\left[ \frac{3\pi}{2G}\int_{a_{3}}^{a_{1}}
\sqrt{V_{0}(a)} da\right]\right.\nonumber\\ &\times&
\left.\cos\left[ \frac{3\pi}{2G}\int_{a}^{a_{3}}\sqrt{ -V_{0}(a)}
da-\frac\pi4\right]\right\} , \, (a <a_{3})\nonumber\\
\label{cg26a}
\\
\Psi(a) &=& C\left\{ \exp\left[ -
\frac{3\pi}{2G}\int_{a}^{a_{1}}\sqrt{ V_{0}(a)}da\right]\right.
\nonumber\\&+& \left. 2i\exp\left[ \frac{3\pi}{2G}\int_{a}^{a_{1}}
\sqrt{
V_{0}(a)}da \right] \right\} , \, (a_{3}< a <a_{1}) \nonumber\\ \label{cg26b} \\
\Psi(a) & = & 2Ce^{i\pi/4}\exp\left[
-i\frac{3\pi}{2G}\int_{a_{1}}^{a}\sqrt{ -V_{0}(a)}da\right] , \,
(a_{1}<a). \nonumber\\ \label{cg26c}
\end{eqnarray}

Having established explicitly the analytical form for the
Hartle-Hawking and tunnelling (Vilenkin) wave functions in
Eq.~(\ref{cg24a})-(\ref{cg24c}) and
Eq.~(\ref{cg26a})-(\ref{cg26c}), respectively, we can now analyze
the transition amplitude \textit{\textbf{A}}, for the universe to
evolve from the classically allowed Lorentzian region between
$0<a<a_{3}$ to the other where $a>a_{1}$ (see figure
\ref{potenciales})\footnote{For a discussion on the general
meaning of tunneling  universes, see, e.g., Ref. \cite{conradi}.}.
This transition amplitude is given by
\begin{equation}
\textit{\textbf{A}}=\exp\left( \mp2\tilde{I}\right)  \label{cg27}
\end{equation}
where the $-$ and $+$ sign correspond, respectively, to the tunnelling
(Vilenkin) and
Hartle-Hawking  wave functions, and we use the expression
\begin{equation}
\tilde I=\frac{3\pi}{2G}\int_{a_{3}}^{a_{1}}\sqrt{\lambda
a(a_{1}-a)(a+a_{1}+a_{3})(a-a_{3})}da.  \label{cg28}
\end{equation}
We henceforth consider the new dimensionless quantity $\lambda\tilde{I}G$
which will be analyzed as function of the parameter
\begin{equation}
X\equiv B^{\frac1{1+\alpha}}\sqrt{\Lambda}  \label{cg29}
\end{equation}
that is also dimensionless. When $X=0$ we recover the well know
closed FRW-DeSitter quantum cosmology case. From Eq.~(\ref{cg28})
it can be seen that $\lambda\tilde{I}G$ decreases when $X$ varies
from $0$ to $2/3$ and hence the transition amplitude,
\textit{\textbf{A}}, for the tunnelling (Vilenkin) boundary
condition increases when $X$ increases (for a fixed value of the
cosmological constant), while for the Hartle-Hawking  boundary
conditions \textit{\textbf{A}} decreases.

\begin{figure}[h]

\includegraphics[width=\columnwidth]{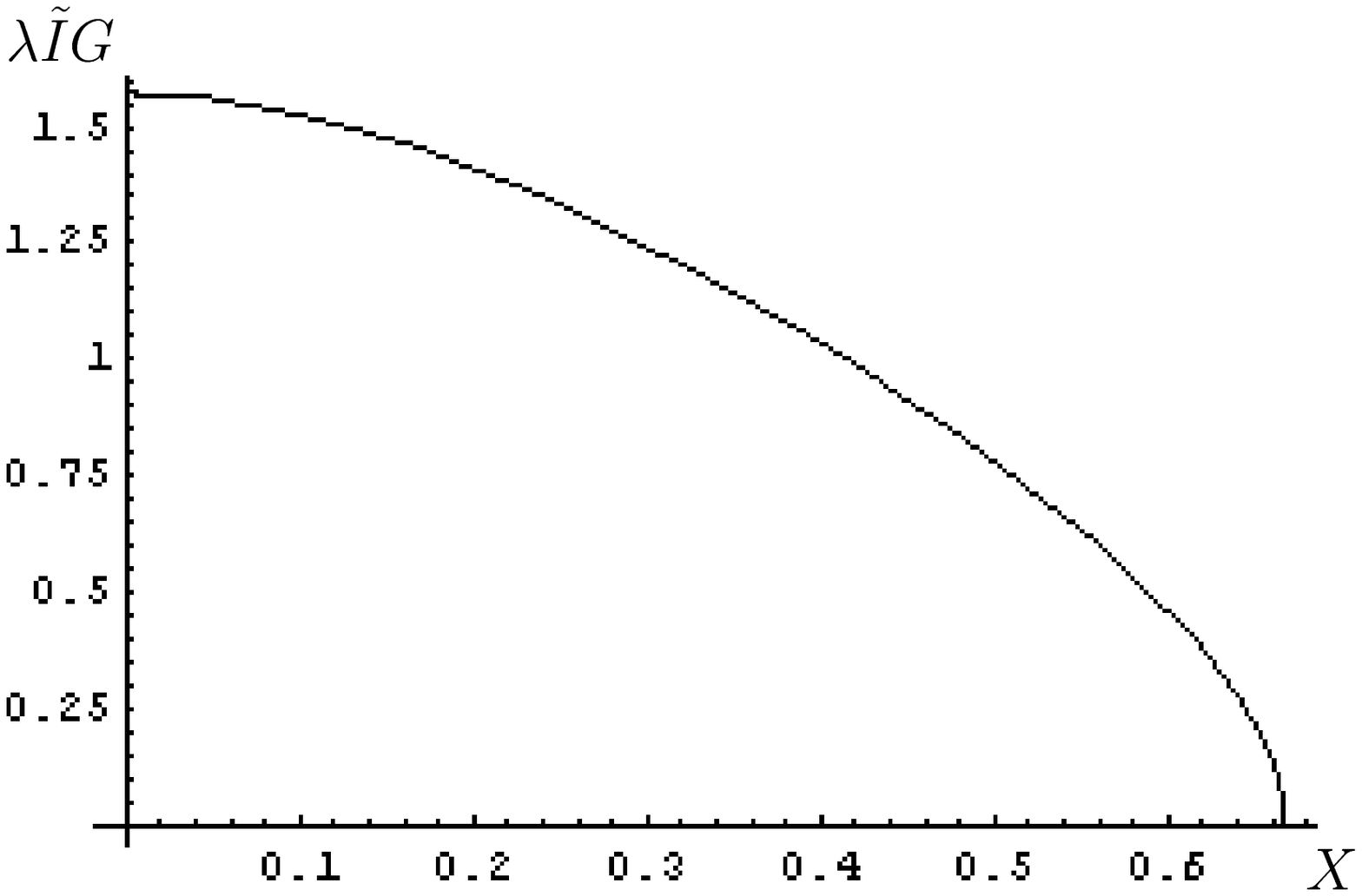}
\caption{This figure shows the behavior of $\lambda \tilde{I} G$,
where $\tilde{I}$ is defined in Eq.~(\ref{cg28}),
in terms of the dimensionless parameter $X$
defined in Eq.~(\ref{cg29}). For $X=0$ we recover the closed
FRW-DeSitter case, while the case $X=2/3$ corresponds to
$\Lambda\tilde{B}^2=4/9$. In this last case there is no tunnelling
between the two Lorentzian regions.}\label{amplitud}

\end{figure}

In terms of the generalized Chaplygin gas parameters $B$ and
$\alpha$, we have that when $\alpha$ increases from $0$ to $1$ and
$B$ is constant and smaller that 1, $X$ also increases and
consequently $\lambda\tilde{I}G$ decreases, which for a fixed
value of $\Lambda$ determines that $\tilde{I}$ also decreases.
This then implies that the transition amplitude for the tunnelling
(Vilenkin) boundary condition increases but the transition
amplitude associated to the no-boundary proposal decreases. The
opposite conclusion is reached when $\alpha$
increases and $B$ is constant and larger than 1. Taking now fixed values of $%
\alpha$ and $\Lambda$, if $B$ increases then $X$ increases. As a
consequence, we conclude that similarly to the previous situation
(for $\alpha$ increasing when $B$ is constant and smaller than 1)
also the transition amplitude for the tunnelling (Vilenkin)
boundary condition increases but the transition amplitude
associated to the Hartle-Hawking  proposal decreases. In
Fig.~\ref{amplitud} we plot how $\lambda \tilde{I}G$ varies as
function of $X$.

\section{Quantization Condition involving  $B$ and $\alpha$}

Let us now establish how the quantum states obtained in the
previous section may determine which sets of values for $B$ and
$\alpha$ are allowed.

In the case of a closed FRW universe with a positive cosmological
constant and a generalized Chaplygin gas, it can be seen that a
singularity is present at $a=0$ as the Ricci scalar diverges. This
divergence in $R$ is present whether we consider the approximation
introduced in Sec.~II [see Eq.~(\ref{cg8})-(\ref{cg10})] or the
the full expression of the energy density of the Chaplygin gas. In
fact, for the latter we have
\begin{eqnarray}
R&=&4\left\{\Lambda+2\pi
G\left[A+\frac{B}{a^{3(1+\alpha)}}\right]^{-\frac{\alpha}{\alpha+1}}
\left[4A+\frac{B}{a^{3(1+\alpha)}}\right]\right\}.\nonumber\\
\label{eqNNew1}
\end{eqnarray}
As it can be shown, for small values of the scale factor, the
Ricci scalar $R$ diverges. Furthermore, for small values of the
scale factor, $R$ in Eq.~(\ref{eqNNew1}) will have  the same
behavior as that we would obtain from the approximations in
Sec.~II. Such divergence could be avoided if
 some other material content dominates the energy density of
the universe at that stage (for example some  effective scalar
field), but we  disregard this possibility in the present work.

The presence of a curvature divergence becomes more complicated in
the FRW  scenario with a generalized Chaplygin gas: the divergence
is present in a Lorentzian (classically allowed) region (see
Ref.~\cite{israel} as well as \cite{5rp,6rp}). A suggestion has
been advanced, designated as DeWitt's argument
\cite{israel,DW1,DW2},
 in that $\Psi(a=0)=0$  is imposed.
The interpretation is that the divergence is not replaced
by an Euclidean conic-singularity-free pole, but instead is neutralized
by making it quantum mechanically inaccessible to wave packets.

Before proceeding, let us also mention the following. Taking
$\lambda=0$ in Eq.~(\ref{cg3}), it can be shown (see Sec.~VI for
more details) that $V(a)$ for $A\ll 1$ is essentially as
represented in Fig.~1 for $V_0 (a)$ with $\lambda \neq 0$ (within
the mentioned approximations). Hence, the validity of the results
presented in this section concerning $B$ and $\alpha$ are not
restricted to our approximations. The physical consequences will
hold in a wider scenario, albeit with some modifications in the
mathematical expressions.

Assuming therefore that the wave function vanishes at the
origin\footnote{See Sec.~VI for a discussion on this issue.},
i.e., $a=0 \Rightarrow \Psi(a=0)=0$, from Eq.~(\ref{cg23}),
(\ref{cg24a}) it is obtained that   in the case of the
Hartle-Hawking wave function we will have
\begin{equation}
\frac{3\pi }{2G}\int_0^{a_3}\sqrt{-V_0(a)}da-\frac \pi 4=n\pi ,\ \
\ n\in \mathbb{Z}. \label{cg25}
\end{equation}
It can be further seen that when $B^{\frac 1{1+\alpha }}$ is small
\begin{equation}
\int_0^{a_3}\sqrt{-V_0(a)}da\approx \int_0^{B^{\frac 1{1+\alpha }}}\sqrt{%
-a^2+aB^{\frac 1{1+\alpha }}}=\frac \pi 8B^{\frac 2{1+\alpha }}.
\label{cgn2}
\end{equation}
From Eq.~(\ref{cg25}) and (\ref{cgn2}) it follows that
\begin{equation}
\frac{3\pi}{4G}B^{\frac 2{1+\alpha}}-1\approx 4n,
\label{cgnn2}
\end{equation}
with  $n\in \mathbb{Z}$.

Recalling that the Vilenkin wave function can be  written  as
\begin{eqnarray}
\Psi(a)&=& C \exp\left[ -\frac{3\pi
}{2G}\int_{a_{3}}^{a_{1}}\sqrt{V_{0}(a)} da\right]\nonumber\\
&\times& \cos\left[ \frac{3\pi}{2G}
\int_{a}^{a_{3}}\sqrt{-V_{0}(a)}
da+\frac\pi4\right]  \nonumber  \\
& + &4iC
 \exp\left[ \frac{3\pi}{2G}\int_{a_{3}}^{a_{1}} \sqrt{V_{0}(a)}
da\right]\nonumber\\&\times&\cos\left[
\frac{3\pi}{2G}\int_{a}^{a_{3}}\sqrt{ -V_{0}(a)}
da-\frac\pi4\right],  \label{cg26bb}
\end{eqnarray}
for the case of $0<a<a_3$ and  assuming again that $\Psi (a=0)=0$,
it now results that
\begin{equation}
\left\{
\begin{array}{c}
\frac{3\pi }{2G}\int_0^{a_3}\sqrt{-V_0(a)}da+\frac \pi 4=\frac \pi
2+k\pi,
\\
\frac{3\pi }{2G}\int_0^{a_3}\sqrt{-V_0(a)}da-\frac \pi 4=\frac \pi 2+\tilde{k%
}\pi,
\end{array}
\right.   \label{cgn3}
\end{equation}
where $k,\tilde{k}\in \mathbb{Z}$. Both conditions in (\ref{cgn3})
cannot generally hold simultaneously. However, there is an
exception. When $\frac{3\pi }{2G}
\int_{a_3}^{a_1}\sqrt{V_0(a)}da$\ $\gg 1$, it is enough that the
last condition holds. This statement can be proven as follows. Let
us take
\begin{eqnarray*}
&&\cos \left[ \frac{3\pi }{2G}\int_0^{a_3}\sqrt{-V_0(a)}da+\frac
\pi 4\right] \nonumber\\ &&+4i\exp \left[ \frac{3\pi
}G\int_{a_3}^{a_1}\sqrt{V_0(a)}da\right]
\times  \\
&&\cos \left[ \frac{3\pi }{2G}\int_0^{a_3}\sqrt{-V_0(a)}da-\frac \pi 4\right]
=0.
\end{eqnarray*}
If we now define
\begin{eqnarray}
%\chi  &\equiv &\exp \left[ \frac{3\pi i}G\int_0^{a_3}\sqrt{-V_0(a)}da\right],
%\label{cgn10} \\
\Upsilon  &\equiv &\exp \left[ \frac{3\pi }G\int_{a_3}^{a_1}\sqrt{V_0(a)}%
da\right],   \label{cgn11}
\end{eqnarray}
we can write
\begin{equation}
\left\{
\begin{array}{c}
\frac{3\pi }G\int_0^{a_3}\sqrt{-V_0(a)}da\ =\frac \pi
2+\tilde{n} \pi,  \\
\sin \left[ \frac{3\pi }G\int_{0}^{a_3}\sqrt{V_0(a)}da\right] =\frac{%
1-4\Upsilon }{1+4\Upsilon },
\end{array}
\right.   \label{cgn12}
\end{equation}
where $\tilde{n}\in \mathbb{Z}.$ These equalities represent
(\ref{cgn3}) and then cannot both hold unless $\Upsilon
\rightarrow 0$ or $\Upsilon \gg 1$. As we are considering the
possibility that the universe tunnels from one Lorentzian region
($a<a_3$) to another Lorentzian region  ($a_1<a$), the parameter
$\Upsilon$ is greater that the unity. Consequently, the case
$\Upsilon \rightarrow 0$ is physically impossible, while the case
$\Upsilon \gg 1$ is admissible. In this situation, we have,  with
$\tilde{n}=2\tilde{k}+1$, that
\begin{equation}
\frac{3\pi }{2G}\int_0^{a_3}\sqrt{-V_0(a)}da=\frac{3\pi }4+\tilde{k}\pi,
\label{cgn13}
\end{equation}
which is equivalent to the second equality in expression
(\ref{cgn3}).

Finally, we can address the question when is the condition
$\frac{3\pi }{2G}\int_{a_3}^{a_1}\sqrt{V_0(a)}da$\ $ \gg 1$
fulfilled. To this aim we write
\begin{equation}
\exp \left[ \frac{3\pi }{2G}\int_{a_3}^{a_1}\sqrt{V_0(a)}da\right] =\exp
\left[ \frac{3\pi }2\tilde{I}\right].   \label{cgn14}
\end{equation}
We know the behavior of $\lambda \tilde{I}G$ in terms of the
parameter $X$ [defined in Eq.~(\ref{cg29})]: $\lambda \tilde{I}G$
will take its largest values when $X$ is small. So in order that
$\exp [ (3\pi/2) \tilde{I}] \gg 1$, it is necessary that $\lambda
\tilde{I}G$ gets its largest values and $\lambda G$ be small,
i.e., $\lambda $ has to be small. Hence, what we have is that the
condition $\Psi (a=0)=0$ applied to the Vilenkin  wave function
implies the universe to be close to a DeSitter model and to have a
small cosmological constant. In this case we obtain that
\begin{equation}
\frac{3\pi }{4G}B^{\frac 2{1+\alpha }}\approx 3+4\tilde{k}.
\label{cgn15}
\end{equation}

\section{Beyond the Hamilton-Jacobi Equation}

In this section we will consider the next order of approximation
in Eq.~(\ref{cg8}), i.e., bringing the terms
$a^{4+3\alpha}(1+\alpha)^{-1}AB^{-\frac{\alpha}{\alpha+1}}$ into
the minisuperspace potential. This will be done
perturbatively\footnote{See Ref.~\cite{qc} and \cite {ck} for a
formal description regarding the expansion of a wave function of
the universe in terms of a perturbative parameter.} with the
parameter $A$ satisfying $A \ll 1$, allowing  a quantum
cosmological analysis beyond the WKB formulation in
Eq.~(\ref{cg5})-(\ref{cg7}). Furthermore, it will allow us to
include explicitly in the analysis a parameter that can be related to the
brane tension in some brane world cosmologies (see
Ref.~\cite{26}).

Let us consider again the Wheeler-DeWitt equation (\ref{cg2}) and
employ the following expansion where we take $A\ll 1$:
\begin{eqnarray}
\Psi (a)=\exp \left[ -\frac 1G\left(
S_0(a)+AS_1(a)+A^2S_2(a)+\cdots \right) \right] .\nonumber \\
\label{cgn17}
\end{eqnarray}
Using the expansion in Eq.~(\ref{cg8}) up to the term in
$a^{4+3\alpha }$ we then obtain in powers of $A$
\begin{eqnarray}
&&-\frac 1{3\pi G}\left[
\frac{dS_0(a)}{da}+A\frac{dS_1(a)}{da}+A^2\frac{
dS_2(a)}{da}\right] ^2\Psi (a)  \nonumber \\
&&-\frac 1{3\pi }\left[
\frac{d^2S_0(a)}{da^2}+A\frac{d^2S_1(a)}{da^2}+A^2
\frac{d^2S_2(a)}{da^2}\right] \Psi (a)  \nonumber \\
&& + \frac{3\pi }{4G}\left[ V_0(a)+AV_1(a)+A^2V_2(a)\right] \Psi
(a)+\cdots  =0,\nonumber\\  \label{cgn18}
\end{eqnarray}
where
\begin{eqnarray}
V_0(a) &=&a^2-\lambda a^4-aB^{\frac 1{1+\alpha }},  \label{cgn19a} \\
V_1(a) &=&-\frac{a^{4+3\alpha }}{2+\alpha }B^{-\frac \alpha {1+\alpha }},
\label{cgn19b}
\end{eqnarray}
with $V_2(a)$ following in a similar manner (but will not be necessary at
this level).

In the WKB approximation where $\frac{d^2S_0(a)}{da^2}$ is neglected, we get
at the $A^0$ and $A^1$ level, respectively, the equations
\begin{eqnarray}
\left( \frac{dS_0(a)}{da}\right) ^2 &=&\frac{9\pi ^2}4V_0(a),  \label{cgn20a}
\\
\frac 2{3\pi G}\frac{dS_0(a)}{da}\frac{dS_1(a)}{da}& - & \frac 1{3\pi }\frac{%
d^2S_1(a)}{da^2}+\frac{3\pi }{4G}V_1(a) =0, \nonumber\\
\label{cgn20b}
\end{eqnarray}
where the former is the Hamilton-Jacobi equation (obtained in
Sec.~II). If we further assume that $S_1(a)$ is slowly varying,
then Eq.~(\ref{cgn20b}) can be approximated by
\begin{equation}
\frac 2{3\pi G}\frac{dS_0(a)}{da}\frac{dS_1(a)}{da}+\frac{3\pi
}{4G} V_1(a)\approx 0.  \label{cgn21}
\end{equation}

If we consider that the turning points are exclusively determined by
the approximate potential $V_0$ and that $V_1$ is a
correction when $A$ is small, then it can be possible to obtain (by
an approximate analytical expression) a quantification rule involving
$\alpha,B$ as well as   $A$. In order to get  analytical expressions for this
case, we should integrate the expression
\begin{equation} \label{NNeq2}
J=\int_{0}^{a_3}\frac{V_1(a)}{\sqrt{-V_0(a)}}da,
\end{equation}
as implied from Eq.~(\ref{cgn20a}) and (\ref{cgn21}). By means of
an approximation similar to the one given in Eq.~(\ref{cgn2}),
namely with  $B^{1/(1+\alpha)}$ being
 small (i.e., $\Lambda\tilde{B}^2$ small),  then $J$ can be approximated
as follows
\begin{equation} \label{NNeq3}
J\approx\int_{0}^{B^{\frac{1}{1+\alpha}}}a^{\frac72+3\alpha}(B^{\frac{1}{1+\alpha}}-a)^{-\frac12}da.
\end{equation}
This  expression can be written analytically as
\begin{equation} \label{NNeq4}
J\approx B^{4+3\alpha}\mathbf{B}\left(9/2+3\alpha,\frac12\right),
\end{equation}
where $\mathbf{B}$ is the beta function (see Ref.~\cite{W,X}).
Quantification rules follow when using
 wave functions in the region $0<a<a_3$ for the two
boundary conditions that we have considered. We will not pursue this issue further in this
paper, as it will be addressed in a future work.

%%%%%%%%%%%%%%%%%%%%%%%%%%%%%%%%%%%%%%%%%%%%%%%%%%%%%%%%%%%%%%%%%%%%%%%%%%%%%%%%%%%%%%%%%%%%%%%%%%%%%%%%%%%%%%%%

Therefore and with suitable choices of boundary conditions we
can then obtain an explicit form for $S_1(a)$ and hence for
\begin{equation}
\Psi (a)\approx \exp \left[ -\frac 1G\left( S_0(a)+AS_1(a)\right).
\right] \label{cgn22}
\end{equation}
 Within this procedure we can thus include explicitly the parameter
$A$ in the wave function of the universe. This may be of
relevance:  a connection between the Chaplygin gas ($\alpha =1$)
and brane-world models is that the brane tension is $\sqrt{A}$
\cite{26}. In the case of a generalized Chaplygin gas the
``generalized'' tension would be $A^{\frac 1{1+\alpha }}$ (see
Ref.~\cite{15}). Therefore, if  $S_1(a)$ is computed then it would
also be possible to extract a quantization
condition involving  $A$ together with $B$ and $%
\alpha $, broadening the scope of Eq.~(\ref{cgnn2}) and
(\ref{cgn15}).

Accepting that the minisuperspace similarities pointed out in
Sec.~II between some brane-world cosmologies and FRW models with
complex scalar fields (inducing the generalized Chaplygin gas
matter content) provide a useful association, then these results
may be of  interest. Of course the range of models herein
considered is yet restricted, but it may direct to potential
insights regarding a quantum mechanical description  of
brane-worlds together with the creation of corresponding
cosmological models.

%%%%%%%%%%%%%%%%%%%%%%%%%%%%%%%%%%%%%%%%%%%%%%%%%%%%%%%%%%%%%%%%%%%%%%%%%%%%%%%%%%%%%%%%%%%%%%%%%%%%%%%%%%%%%%%%
%%%%%%%%%%%%%%%%%%%%%%%%%%%%%%%%%%%%%%%%%%%%%%%%%%%%%%%%%%%%%%%%%%%%%%%%%%%%%%%%%%%%%%%%%%%%%%%%%%%%%%%%%%%%%%%%
%%%%%%%%%%%%%%%%%%%%%%%%%%%%%%%%%%%%%%%%%%%%%%%%%%%%%%%%%%%%%%%%%%%%%%%%%%%%%%%%%%%%%%%%%%%%%%%%%%%%%%%%%%%%%%%%
%%%%%%%%%%%%%%%%%%%%%%%%%%%%%%%%%%%%%%%%%%%%%%%%%%%%%%%%%%%%%%%%%%%%%%%%%%%%%%%%%%%%%%%%%%%%%%%%%%%%%%%%%%%%%%%%
%%%%%%%%%%%%%%%%%%%%%%%%%%%%%%%%%%%%%%%%%%%%%%%%%%%%%%%%%%%%%%%%%%%%%%%%%%%%%%%%%%%%%%%%%%%%%%%%%%%%%%%%%%%%%%%%
%%%%%%%%%%%%%%%%%%%%%%%%%%%%%%%%%%%%%%%%%%%%%%%%%%%%%%%%%%%%%%%%%%%%%%%%%%%%%%%%%%%%%%%%%%%%%%%%%%%%%%%%%%%%%%%%

\section{Conclusions and Discussions}

As a summary, let us point the main results conveyed in this
paper. Our purpose was twofold.

\vspace{0.5cm}

First, it was our aim to  present a quantum mechanical description of
a closed FRW model with a positive cosmological constant and a
generalized Chaplygin gas. The physical scenario we are
contemplating is that, underlying a classical universe, there was
a wave function vestige (i.e., a rapidly oscillating state with a
particularly small amplitude)
whose physical implications would
emerge within the epoch of cosmological dust dominance.  Such
specific
remnant component of the original wave function of the universe
had been present during that period of the  universe evolution. It can be
interpreted as a robust component (with respect to
decoherence) in the sense introduced in Ref.~\cite{CKZeh,Decoh,Diosi}.
In other words, after the universe reached a dust stage, the transition towards an
accelerating expansion would be associated to a quantum mechanical
transition for that specific wave function remnant component of
the original state. In this context, we obtained the Hamiltonian
and the corresponding Wheeler-DeWitt equation associated with such
transition. In particular, we established a range for the
variables and parameters of the model where a linear approximation
for the effective minisuperspace potential could be used.

The quantum mechanical amplitudes for the transitions from a
 dust stage towards an expanding accelerating phase were
  obtained  either in the form of a Hartle-Hawking  or Vilenkin wave
function. We established how these transition amplitudes vary
according to the allowed values for $\alpha$ and $B$:
when either
  $\alpha$
or  $B $ increase, the Vilenkin state (generally, see Sec.~III)
allowed the probability for the universe to tunnel towards the
accelerated stage to increase, while the opposite situation
occurred for the Hartle-Hawking wave function.

Subsequently, we  investigated whether this  quantum mechanical
formulation for the transition from a dust stage towards an
exponentially accelerated phase determined any physical
consequences regarding the Chaplygin gas cosmological scenario. We
then found that the Hartle-Hawking and Vilenkin  wave functions
could imply  that the parameters $\alpha$ and $B$ would be
related and restricted to a
quantization condition\footnote{Let us mention that, in a different
physical context, other quantization conditions (bearing formal similarities)
were obtained in ref.\cite{CKother}, when requiring normalization
of the quantum states.} characterized by the presence of an integer $n$.

This result is obtained as we impose the condition $\Psi(a=0)=0$
corresponding to the presence of a divergence in the curvature.
The divergence is generic, i.e., it is not a consequence of our
linear approximation. The presence of such divergence in a
classically allowed region can be dealt following, e.g., the
approach suggest in
 Ref.~\cite{israel,5rp,6rp} (also designated as DeWitt's argument): the divergence is not
 replaced by an Euclidean conic-singulatity-free
 pole, being neutralized by making it quantum mechanically inaccessible.
 Let us remember that, e.g., the tunneling (Vilenkin) boundary condition was thought of for a
 problem of (semi-classical) creation of the universe, in analogy with
 the tunneling process in quantum mechanics. Note, however, that
 the singularity in our model is located in a classical region, before
 tunneling. We can consider waves proceeding from a classical
 region, with a probability to  exit  from it
 towards the other classical region (see figure 1).

 Nevertheless\footnote{The authors are grateful to C. Kiefer and A. Vilenkin for
 their comments on this issue.}, it is only a sufficient but not necessary
 that the wave function vanishes at the origin in order to avoid singularities.
 On the one hand, both the Hartle-Hawking and Vilenkin wave functions can be taken generally
 as non-zero at $a=0$
 (being a minimum (Hartle-Hawking) or a Green's function (Vilenkin)), increasing or decreasing
 through the barrier. The singularity could be smoothed by means of some non-trivial measure in
 the inner product.
 On the other hand,  quantum mechanical situations
 with singularities occur and are dealt with, as with the
 solution of the Dirac equation for the ground state
 of the hydrogen atom which is singular when $r\rightarrow 0$.
 We have, however, herein chosen  to follow the approach presented in  Ref.~\cite{israel,5rp,6rp}
 to address the transition from a dust phase towards an accelerating expansion stage
 by means of a quantum mechanical tunneling mediating the evolution
 from an initial  Lorentzian (classically allowed) region  to the current
 accelerating epoch.

 \vspace{0.5cm}

Second, it was also our objective to introduce
a discussion (and promote subsequent research) on the
interesting similarities between
the quantum cosmology of a specific brane-world models
\cite{israel} and  FRW quantum cosmologies with a Chaplygin gas
content. This was based on the feature that  FRW models with a
Chaplygin gas may constitute a phenomelogical tool to investigate
brane-world models: both frameworks share a physical minisuperspace where
 two Lorentzian sectors are separated
by a classically forbidden Euclidean region \cite{5rp,6rp}.
This suggests a view
 that FRW models with a generalized Chaplygin gas content
may enable us  to investigate specific aspects within
brane-worlds. E.g., the quantum cosmological creation of the
universe, by means of an Hamiltonian formulation, taking into
account the presence of potential symmetries and conserved charges
that could be related with physical quantitites in brane-world
physics \cite{ekpy,israel,leon,odintsov,j,k,l,m}.

 This
association between FRW cosmologies with a generalized Chaplygin
gas and some brane-world cosmologies is strengthen  when
considering the next order of approximation in Eq.~(\ref{cg2}) and
(\ref{cg10}) (i.e., going beyond the Hamilton-Jacobi equation
(\ref{cg7})): see Eq.~(\ref{cgn17}) - (\ref{cgn22}). This  order
of approximation was  described through of a perturbative
expansion in the parameter $A$ present in the Chaplygin gas energy
density. Let us remind that a connection between the Chaplygin gas
($\alpha =1$) and brane-world models is that the brane tension is
$\sqrt{A}$ \cite{26}. For the  case of a generalized Chaplygin gas
the ``generalized'' tension would be $A^{\frac 1{1+\alpha }}$. A
subsequent
 quantization condition involving  $A$ as well as  $\alpha$ and
$B$  would be retrieved, depending on the boundary condition
(Hartle-Hawking or Vilenkin) chosen.

\begin{figure}[h]
\includegraphics[width=5cm,angle=90]{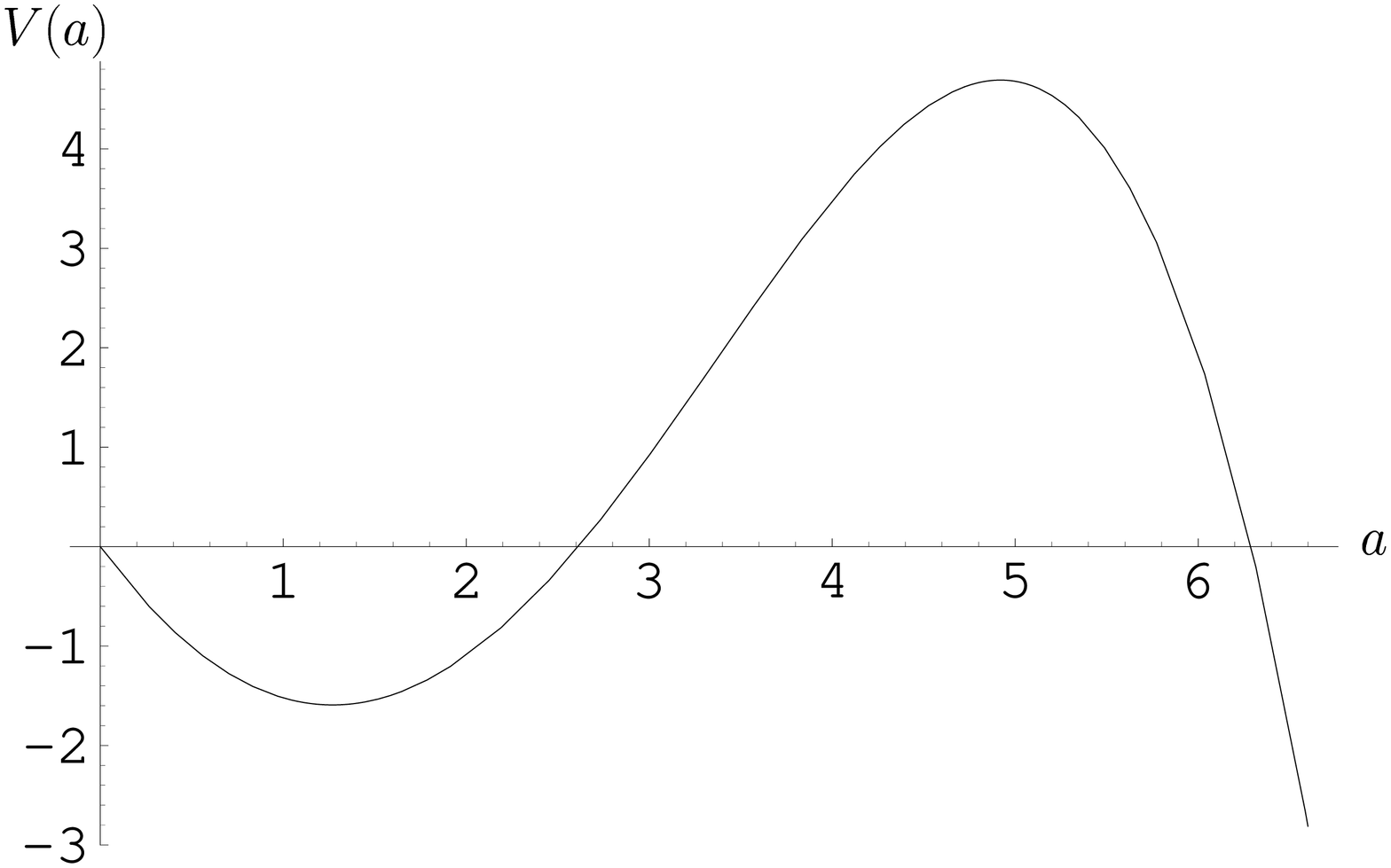}
\caption{This figure shows the behavior of the potential $V(a)$,
when the cosmological constant, $\lambda$, is zero and the
constant $A$ is very small. The physical situation described in
this case is similar to the one in Fig.~\ref{potenciales}, i.e.
two classically allowed  regions disconnected and a Euclidean
one.}\label{comparison}
\end{figure}

Let us now discuss the presence of the positive cosmological
constant, $\Lambda$, in our minisuperspace scenario. It may seem
that its purpose is merely to induce the tunnelling towards a
DeSitter classical stage, as Eq.~(\ref{cg10}) suggests.
Nevertheless, it is worthy to point the following. If we take
$\lambda=0$ in Eq.~(\ref{cg3}) and depict $V(a)$ for different
values of $\alpha$ and with $A\ll 1$, we get a physical situation
as described in Fig.~\ref{potenciales} (see Fig.~\ref{comparison}
for a comparison),
where the behavior of an effective cosmological constant is
induced through the parameter $A$. However, the analytical quantum
treatment of a minisuperspace with $V(a)$ in Eq.~(\ref{cg3}) with
generic values of $A, B$ and $\alpha$ is rather complex. The minisuperspace model
described within this paper with positive cosmological constant,
$\Lambda$, is analytically feasible, providing  a similar minisuperspace dynamics through
Eq.~(\ref{cg10}). The presence of $\lambda$ (pertinent for larger
values of $a$) may be interpreted as the lowest order effect of
the effective cosmological constant behavior induced in the
generalized Chaplygin gas scenario. This provides explicit
analytical expressions for the transition amplitudes
 (from  an initial  Lorentzian (classically allowed) region  to the current
 accelerating epoch through a tunneling process)
to be computed. Improved expressions could be considered by taking
next orders of approximation, beyond the Hamilton-Jacobi
Eq.~(\ref{cg7}), as indicated in Sec.~V.

Another issue that is important to mention is the following. In
Eq.~(\ref{cg5}) we have a superposition of states, that may be considered
from the point of view of decoherence (see
Ref.~\cite{Decoh} for an extensive and thorough review). In
generic quantum cosmology, a full wave function could take the
form $\Psi = \Sigma_{(n)} C_{(n)} e^{iS_{0}^{(n)}} \chi^{(n)}$
where the $S_{0}^{(n)}$ are solutions to the Hamilton-Jacobi
equation (treating the gravitational part) and the states
$\chi^{(n)}$ are matter states obeying $i\bigtriangledown
S_{0}^{(n)} \bigtriangledown \chi^{(n)} = H^{(n)}\chi^{(n)}$,
where $H^{(n)}$ is the corresponding matter Hamiltonian (see
Ref.~\cite {Decoh}). Such expansion for $\Psi$ has some relevant
non-classical features. On the one hand, it is a superposition of
many WKB states. Only if those different branches are dynamically
independent of each other we recover a background spacetime
without quantum mechanical interferences. On the other hand,
restricting to a single component in $\Psi$  may not yet mean a
classical space-time as the WKB function may spread over all
configuration space. A decoherence process is needed thus prior to
identify a classical behavior.

Concerning our case study, we focused our attention in the
transition amplitudes for the universe to proceed from a classical
initial region,  through a quantum mechanical tunneling sector,
towards another classical region where an accelerating stage takes
place. We are taking that different WKB components will decohere
as discussed in \cite{Decoh} and considering solely a robust
component  (with respect to decoherence, in the sense
introduced in Ref.~\cite{CKZeh,Decoh,Diosi}) of the original wave
function. This component would induce a  transition from a dust dominating
stage towards an accelerated epoch as described in sections I and
II. A discussion on decoherence issues in our model (or other models
where the Euclidean (classically forbidden) region contour is
generically a closed curve) may be the subject of future research
work.

Finally, it may be of interest to indicate additional  possible
subsequent directions of research to explore. An interesting
option is to investigate whether other brane-world models (see,
e.g. Ref.~\cite{ekpy,israel,leon,odintsov,j,k,l,m}) can admit a
phenomenological description in terms of a generalized Chaplygin
gas minisuperspace cosmology. This could be investigated by means
of suitable canonical transformations at the Hamiltonian
formulation. Another possibility is to pursue the analysis
of the next orders of approximation
(beyond the
Hamilton-Jacobi equation as in section V)
in a broader
perspective  \cite{qc,ck}. This would mean extracting
the corresponding functional Schrodinger equation from the
Wheeler-DeWitt equation, together with quantum gravitational
corrections to it (see Ref.~\cite{ck} for more details). Moreover,
the quantum cosmology of FRW models with a generalized Chaplygin
gas could be investigated from the following perspective: taking
into account the Chaplygin gas influence not as as additional term
in the superpotential depending on $a$, but as a couple of
independent conjugate dynamical variables, which would imply a
modification of the Wheeler-DeWitt equation. The  guidelines of Ref.~\cite{Jackiw}
could be of importance towards such Hamiltonian formulation.
This will bring changes as the dimension of the configuration
space will increase and the hyperbolic nature of the kinetic term
has to be considered.

\appendix

\section{Positive Roots in case $\Lambda\tilde{B}^2 <
4/9$}

Let us write $V_0(a)$ as
\begin{equation}
V_0(a)=-\lambda af(a)  \label{cga1}
\end{equation}
with
\begin{equation}
f(a)=a^3+b_2a^2+b_1a^1+b_0a^0,  \label{cga2}
\end{equation}
where $b_2=0$, $b_1=-\lambda ^{-1}$, $b_0=\frac{\tilde{B}}\lambda
$. Using Ref.~\cite{X}, we define
\begin{equation}
q=-\Lambda ^{-1},\ \ r=-\frac 32\frac{\tilde{B}}\Lambda \Rightarrow
q^3+r^2=-\Lambda ^{-3}+\frac 94\frac{\tilde{B}^2}{\Lambda ^2},  \label{cga3}
\end{equation}
and  then obtain for the case $q^3+r^2<0$, i.e., $\Lambda\tilde{B}^2
<\frac 49$, the following roots for $f(a)=0$:
\begin{eqnarray}
a_1 &=&2\left| x\right| ^{\frac 13}\cos \left( \frac \theta 3\right),
\label{ccga4a} \\
a_2 &=&-2\left| x\right| ^{\frac 13}\cos \left( \frac{\theta}{3}
-\frac{\pi}{3}\right),
\label{cga4b} \\
a_3 &=&-2\left| x\right| ^{\frac 13}\sin \left( \frac \pi 6-\frac
\theta 3\right).   \label{cga4c}
\end{eqnarray}
In the last expression $\left| x\right| =\Lambda ^{-\frac 32}$ and
$\tan \theta =-\frac{\sqrt{1-\frac 94\tilde{B}^2\Lambda }}{\frac
32\tilde{B}\sqrt{\Lambda }}$. Moreover, we have that $a_1-a_3>0$,
together with
\begin{equation}
a_1>a_3>0>a_2.  \label{cga5}
\end{equation}
Therefore, $V_0(a)$ has three real roots determining two
Lorentzian regions and one classically forbidden region for the
range of $a\geq 0$.

\vspace{.5cm}

\acknowledgments

MBL thanks the support of a CENTRA-IST BPD fellowship for the
duration of her visit at the Universidade da Beira Interior as
well as for kind hospitality and a warm atmosphere. MBL is
supported by the Spanish Ministry of Education, Culture and Sport
(MECD). MBL is also partly supported by DGICYT under Research
Project BMF2002 03758. The authors are grateful to C. Barcelo, D.
Clancy, J. Garriga, P.F. Gonz\'{a}lez-D\'{i}az, A. Yu.
Kamenshchik, C. Kiefer, J. Lidsey, R. Maartens, A. Vilenkin, D.
Wands and A. Zhuk for useful comments. Finally, the authors thank
R. Lazkzos for the invitation to participate in the Brane
Cosmology Workshop, Bilbao, Spain, where some issues introduced in
this paper were discussed.

\end{document}